\documentclass{article}
\font\fthreei=cmti10 scaled\magstep1

\font\fthreebb=cmbx10 scaled\magstep2

\def\a{^{\underline a}}

\begin{document}
\centerline {\fthreebb Causa Efficiens versus Causa Formalis:}
\vskip 0.09 true cm
\centerline {\fthreebb origens da discuss\~ao moderna sobre a}
\vskip 0.09 true cm
\centerline {\fthreebb dimensionalidade do espa\c{c}o\footnote{Trabalho
apresentado por R. Moreira Xavier no VIII Col\'oquio de Hist\'oria da
Ci\^encia --  \textit{Espa\c{c}o e Tempo} --
realizado em \'Aguas de Lind\'oia, S\~ao Paulo, de 14 a 17 de outubro de 1993.}}

\vskip 1. cm
\centerline {\fthreei F. Caruso \& R. Moreira Xavier}
\bigskip
\smallskip
\centerline {Centro Brasileiro de Pesquisas F\'{\i}sicas}
\centerline {Rua Dr. Xavier Sigaud 150, 22290-180, Rio de Janeiro, RJ, Brazil}

\vskip 1. cm

\centerline {\textbf{Resumo}}
\vskip 0.3 true cm

\noindent Discute--se a rela\c{c}\~{a}o entre os crit\'erios metacient\'{\i}ficos
utilizados para explicar, ou impor limites sobre a dimensionalidade do
espa\c{c}o f\'{\i}sico e os sistemas de explica\c{c}\~{a}o causal dominantes nos
cor\-res\-pondentes per\'{\i}odos hist\'oricos.  Examinam--se as importantes
contribui\c{c}\~{o}es de Arist\'oteles, Kant e Ehrenfest ao problema da
dimensionalidade, as quais se ap\'oiam em explica\c{c}\~{o}es causais distintas: em
Arist\'oteles, {\it causa materialis}, no jovem Kant, \textit{causa efficiens}
e, em  Ehrenfest, uma engenhosa combina\c{c}\~{a}o de \textit{causa efficiens} e
{\it causa formalis}.  Enfatiza--se a crescente valoriza\c{c}\~{a}o da
\textit{causa formalis} nas abordagens f\'{\i}sicas contempor\^aneas deste problema.
\vskip 0.3 true cm

\centerline {\textbf{Abstract}}
\vskip 0.3 true cm

\noindent Metascientific criteria used for explaining or constraining
physical space dimensionality and their historical relationship to
prevailing causal systems are discussed.  The important contributions by
Aristotle, Kant and Ehrenfest to the dimensionality of space problem are
considered and shown to be grounded on different causal explanations: {\it
causa materialis} for Aristotle, {\it causa efficiens} for young Kant and
an ingenious combination of {\it causa efficiens} and {\it causa formalis} for
Ehrenfest. The prominent and growing r\^ole played by {\it causa
formalis} in modern physical approaches to this problem is emphasized.

\vspace*{0.4cm}
\vspace*{0.4cm}

\newpage

\section{Introdu\c{c}\~{a}o}
\bigskip

          Este trabalho situa-se na conflu\^encia de dois problemas. Por um lado, a quest\~ao da dimensionalidade do espa\c{c}o f\'{\i}sico, cuja
discuss\~ao pode ser desenvolvida, de um ponto de vista moderno, a partir da pergunta de Ehrenfest:

\begin{quotation}
{\noindent \it ``Qual o papel desempenhado pela tridimensionalidade do espa\c{c}o nas leis fundamentais da F\'{\i}sica?}'' (EHRENFEST, 1920, p.~440);
\end{quotation}

\noindent por outro lado, o problema da interdepend\^encia entre os
conceitos de causa, utilizados pela F\'{\i}sica num determinado momento
hist\'orico, e o est\'agio de evolu\c{c}\~{a}o da pr\'opria F\'{\i}sica.

          Trata-se aqui de discutir os esquemas explicativos (causais) que
est\~ao por tr\'as de diversas tentativas de tirar a {\it
tridimensionalidade} do territ\'orio dos {\it dados iniciais} (mat\'eria,
extens\~ao e espa\c{c}o) e coloc\'a-la no universo de problemas da F\'{\i}sica.
Quest\'{o}es hist\'oricas e epistemol\'ogicas relacionadas \`as obras de
tr\^es autores fundamentais --- Arist\'oteles, Kant e Ehrenfest --- ser\~ao
tratadas em certo detalhe. Pretende--se examinar o que eles dizem sobre a
dimensionalidade do espa\c{c}o e, em seguida, discutir as concep\c{c}\~{o}es de
causa subjacentes \`as an\'alises destes autores. Antes, entretanto, \'{e}
preciso situar o problema da dimensionalidade do espa\c{c}o, em seus
aspectos gerais.
\bigskip

\section{A problem\'atica da dimensionalidade do espa\-\c{c}o:
F\'{\i}sica e Matem\'{a}tica}
\bigskip

         Whitrow chama a aten\c{c}\~{a}o de que o problema da
dimensionalidade do espa\c{c}o a\-pre\-senta um car\'ater dual,
envolvendo a F\'{\i}sica e a Matem\'atica (WHITROW, 1955, pp. 13-31).
Segundo ele, primeiro \'e necess\'ario que se questione o significado
de um espa\c{c}o ter um certo n\'umero de dimens\~{o}es --- isto diz
respeito ao dom\'{\i}nio da Matem\'atica.  Depois, segue-se a quest\~ao
de por que este n\'umero \'e precisamente 3 --- a este segundo ponto,
espera-se, que a F\'{\i}sica possa contribuir, de forma significativa,
elaborando um co\-nhecimento mais profundo das {\it peculiaridades
que distinguem o espa\c{c}o tridimensional} de outros, postas em
destaque na pergunta de Weyl:

\begin{quotation}
{\noindent \it ``... se Deus, ao criar o Mundo, decidiu fazer o espa\c{c}o
tridimensional, pode--se chegar a uma explica\c{c}\~{a}o `razo\'avel' deste fato,
desvelando tais peculiaridades?''} (WEYL, 1949).
\end{quotation}

      Ora, claro est\'a que, desde a revolu\c{c}\~{a}o galileana, passa a existir
uma not\'avel interdepend\^encia entre a F\'{\i}sica e a Matem\'atica na
descri\c{c}\~{a}o da Natureza, passando esta \'ultima a ser vista cada vez mais
como a linguagem adequada \`a F\'{\i}sica. No entanto, do ponto de vista
l\'ogico, a Matem\'atica tem suas limita\c{c}\~{o}es intr\'{\i}nsecas: quando se
constroem conceitos  fundamentais para qualquer Teoria F\'{\i}sica -- como o
de {\it espa\c{c}o f\'{\i}sico} -- a partir de conceitos matematicamente bem
definidos -- como o de {\it espa\c{c}o geom\'etrico} --, torna-se
extremamente complexo e intricado explicitar os efeitos dessas limita\c{c}\~{o}es,
inerentes \`a Matem\'atica, sobre a F\'{\i}sica (SCHENBERG, 1985; COSTA \&
DORIA, 1991; BARROS, 1991), em particular, seus reflexos sobre as
qualidades espec\'{\i}ficas do espa\c{c}o f\'{\i}sico, como a dimensionalidade. {\it
Ipso facto}, a dualidade a que se refere Whitrow parece--nos hoje
injustificada do ponto de vista epistemol\'{o}gico, embora esteja na origem
hist\'orica da abordagem moderna do problema da dimensionalidade (JAMMER,
1954). Al\'em disto, injustificada, por e\-xem\-plo, quando se
considera a profunda rela\c{c}\~{a}o entre Geometria e F\'{\i}sica, contida no
programa de Einstein. Como observou Jammer:
\begin{quotation}
{\noindent \it ``Foi Einstein quem esclareceu como a geometria (...) cessa
de ser uma ci\^encia axiom\'atico-dedutiva e torna-se uma entre as
ci\^encias naturais; a mais velha de todas, na verdade.}'' (JAMMER, 1954, p.~170).
\end{quotation}

    Toda essa dificuldade que permeia a F\'{\i}sica Contempor\^anea resume-se em
outro coment\'ario de Jammer:
\begin{quotation}
{\noindent \it (...) a estrutura do espa\c{c}o f\'{\i}sico n\~ao \'e, em \'ultima
an\'alise, nada de dado na natureza ou de independente do pensamento
humano. \'E uma fun\c{c}\~ao do nosso esquema conceitual.}'' (JAMMER, 1954,
p.~171).
\end{quotation}

          Para se ir al\'em na discuss\~ao deste problema seria
necess\'ario indagar at\'e onde a Matem\'atica e, em particular, o {\it
espa\c{c}o cont\'{\i}nuo} s\~ao necess\'arios e adequados \`a descri\c{c}\~{a}o dos
fen\^omenos f\'{\i}sicos, quest\~ao esta que transcende o escopo deste ensaio.
Basta mencionar  que continuaremos a nos guiar pelos paradigmas
mate\-m\'a\-ticos vigentes, pelo menos enquanto n\~ao tivermos que
enfrentar na nossa concep\c{c}\~{a}o de espa\c{c}o--tempo f\'{\i}sico os reflexos das
limita\c{c}\~{o}es, mencionadas acima, inerentes a todos os sistemas axiomatizados
como a Matem\'atica.  Pretendemos aqui t\~ao somente mostrar que o enfoque
dado ao problema da dimensionalidade do espa\c{c}o depende fortemente da
concep\c{c}\~{a}o de {\it causa}\footnote{A palavra \textit{causa} \'e usada \textit{lato sensu}, \textit{i.e.}, relacionada \`{a} ideia geral de explica\c{c}\~{a}o \textit{Cf.} Sec.~(III).} dominante num certo per\'{\i}odo hist\'orico, concep\c{c}\~{a}o esta que, na Ci\^encia Moderna, est\'a intimamente ligada, como veremos, ao
est\'agio de desenvolvimento da Matem\'atica.

          Veremos, na pr\'oxima Se\c{c}\~{a}o, ainda que de modo bastante
esquem\'atico, o segundo ponto a que nos referimos acima, {\it i.e.}, o
impacto dos diversos conceitos de causa sobre a evolu\c{c}\~{a}o da F\'{\i}sica e
vice-versa. \`A luz destas considera\c{c}\~{o}es, discutiremos nas Se\c{c}\~{o}es IV, V
e VI, respectivamente, as contribui\c{c}\~{o}es de Arist\'oteles, Kant e
Ehrenfest \`a quest\~ao da dimensionalidade. Algumas considera\c{c}\~{o}es finais
ser\~ao apresentadas na Se\c{c}\~{a}o VII.

\section{Das quatro causas \`{a}s quatro causas}
\bigskip

          O sistema explicativo de Arist\'oteles, baseado nas quatro
causas, discutido, ela\-bo\-rado e desenvolvido pelos escol\'asticos ({\it
causa materialis, formalis, efficiens} e {\it finalis}), vai ser adotado
pela cultura ocidental at\'e o Renascimento (BUNGE, 1979, p. 32; KUHN,
1977, pp. 51-61), pelo menos.  Em Galileu, a {\it causa formalis} passa a
ter destaque (MOREIRA XAVIER, 1986), devido ao papel que a geometria tem em
seu sistema (KOYR\'E, 1966). Com o surgimento do programa cartesiano de
pesquisa, as explica\c{c}\~{o}es formais e finais s\~ao abandonadas e,
a partir dessa \'epoca, a {\it causa efficiens} vai, pouco a pouco,
ocupando o lugar central dos esquemas explicativos da F\'{\i}sica pelos motivos
discutidos em (BUNGE, 1979; KUHN, 1977; MOREIRA XAVIER, 1986).  \'E claro
que {\it causa efficiens}, agora, adquire um sentido diferente do que se
encontra em Arist\'{o}teles. J\'a se delineia a vis\~ao mecanicista do Mundo
(DIJKSTERHUIS, 1959) e {\it causa efficiens} passa a ser, antes de tudo,
a\c{c}\~{a}o ({\it actio}), ou seja, ``for\c{c}a''\footnote{Note que \textit{for\c{c}a} aqui n\~{a}o tem o significado moderno, mas sim o da \'{e}poca que, visto com os
olhos de hoje, tem muito de \textit{energia, momentum, etc.}
(JAMMER, 1957).}

   Para os newtonianos o programa de pesquisa gira em torno da
determina\c{c}\~{a}o das for\c{c}as que geram os movimentos. Esquematicamente,
podemos dizer que este programa, originado em Descartes, ganha corpo em
Newton, \'e formalizado por Euler, e culmina em
Laplace\footnote{O leitor
interessado em maiores detalhes sobre a evolu\c{c}\~{a}o do conceito de causa na
f\'{\i}sica p\'os-newtoniana pode reportar-se, p. ex., a (MOREIRA XAVIER,
1986).}

\begin{quotation}
{\noindent \it ``N\'os devemos considerar o estado presente do Universo
como efeito de seu estado anterior, e causa do que se deve seguir. Uma
Intelig\^encia que, por um dado instante, conhecesse todas as for\c{c}as de
que a natureza \'e animada e a situa\c{c}\~{a}o respectiva dos seres que a
comp\~oem, se fosse suficientemente vasta para submeter esses dados ao
c\'alculo, abra\c{c}aria na mesma f\'ormula os movimentos dos maiores corpos
do universo e os do \'atomo mais leve: nada seria incerto para ela e o
futuro, como o passado, estaria presente aos seus olhos}'' (LAPLACE, 1814).
\end{quotation}

\noindent \'E o predom\'{\i}nio absoluto da {\it causa efficiens} e do determinismo mecanicista.

A vis\~ao atomista da mat\'eria vai se fortalecer nesse ambiente
cultural, a ponto de numa confer\^encia intitulada ``Os Confins do
Conhecimento da Natureza",  em 1880, o fisi\'ologo du Bois-Reymond
afirmar que a autenticidade de uma ci\^encia estaria, sobretudo, na
sua fundamenta\c{c}\~{a}o na mec\^anica dos \'atomos:

\begin{quotation}
\noindent  ``\textit{Se imagin\'assemos todas as transforma\c{c}\~{o}es do mundo material
resolvidas em movimentos de \'atomos, produzidos por uma for\c{c}a central
cons\-tan\-te, o universo seria cientificamente conhecido. O estado do
mundo durante um diferencial de tempo apareceria como imediato efeito de
seu estado durante o diferencial de tempo precedente, e como causa direta
do seu estado durante o diferencial de tempo sucessivo. \rm{Lei} e
\rm{acaso} seriam somente \rm {diferentes} \rm {nomes}
\rm{da} \rm{necessidade mec\^anica.}}\footnote{Os grifos s\~ao
nossos.}'' (du BOIS--REYMOND, 1891, p.~18)
\end{quotation}

     Por outro lado, \'e na descri\c{c}\~{a}o do calor, visto como algo que
se propaga no cont\'{\i}nuo, que vamos encontrar a origem de um novo estilo de
fazer Ci\^encia: Fourier preocupa-se em descrever o modo pelo qual o calor
se propaga -- atrav\'es de {\it leis simples e cons\-tan\-tes} -- sem
discutir a ess\^encia do calor -- as suas {\it causas prim\'arias} --
como se depreende do Discurso Preliminar da {\it Teoria Anal\'{\i}tica do
Calor}:
\begin{quotation}
{\noindent \it ``As \underbar {causas prim\'arias} nos s\~ao
desconhecidas, mas est\~ao sujeitas a \underbar {leis} \underbar
{simples} \underbar {e constantes}, que podem ser descobertas pela
observa\c{c}\~{a}o, cujo estudo constitui o objeto da filosofia natural. O
calor, como a gravidade, penetra todas as subst\^ancias do Universo,
seus raios ocupam todas as partes do espa\c{c}o. O objetivo de nosso
trabalho \'e estabelecer as leis matem\'aticas a que este elemento
obedece. A teoria do calor, daqui em diante, cons\-ti\-tuir\'a um dos
ramos mais importantes da f\'{\i}sica geral (...). \underbar {Qualquer}
\underbar {que} \underbar {seja} \underbar {o} \underbar {\^ambito}
\underbar {das} \underbar {teorias} \underbar {mec\^anicas},
\underbar {elas} \underbar {n\~ao} \underbar {se} \underbar {aplicam}
\underbar {aos} \underbar {efeitos} \underbar {do} \underbar {calor}.
Estes constituem um tipo especial de fen\^omeno, e n\~ao podem ser
explicados pelos princ\'{\i}pios do movimento e do equil\'{\i}brio.$^{(5)}$}''
(FOURIER, 1822).
\end{quotation}

       Mas note que, de certa forma, Newton tamb\'em faz algo semelhante:
na sua {\it Opticks} (NEWTON, 1730, p. 400) ele admite a exist\^encia
dos \'atomos e procura, nos {\it Principia}  (NEWTON, 1726), descrever as
intera\c{c}\~{o}es da mat\'eria e n\~ao explicar suas origens. Tanto em Newton
quanto em Fourier h\'a claramente um deslocamento da pergunta do {\it
porqu\^e} ao {\it como}. Sendo assim em que, ent\~ao, os dois
programas v\~ao diferir? \'E justamente a introdu\c{c}\~{a}o de um fluido
imponder\'avel e sutil -- o cal\'orico -- que vai fazer a diferen\c{c}a: a propaga\c{c}\~{a}o de uma subst\^ancia fluida no espa\c{c}o cont\'{\i}nuo
ir\'a envolver varia\c{c}\~{o}es de certa grandeza no espa\c{c}o e no tempo,
e, al\'em disto, as coordenadas espaciais passam a ser tamb\'em um
par\^ametro como o tempo: isto implicar\'a no uso de equa\c{c}\~{o}es
diferenciais parciais. O enfoque do problema \'e assim desviado para
a busca de uma equa\c{c}\~{a}o diferencial que descreva o fen\^omeno
f\'{\i}sico, ou seja, para a busca da {\it forma}. Em outras palavras, a
equa\c{c}\~{a}o diferencial \'e a {\it causa formalis} do fen\^omeno em
quest\~ao. Note que a f\'{\i}sica dos fluidos imponder\'aveis e sutis
(fluido el\'etrico, cal\'orico, {\it etc.})  representou uma certa
desmaterializa\c{c}\~{a}o das explica\c{c}\~{o}es, que preparou o terreno para a
introdu\c{c}\~{a}o de conceitos como linhas de for\c{c}a, no caso
el\'etrico, e em \'ultima inst\^ancia, do conceito de {\it campo}.
Isto marca o retorno da {\it causa formalis} ao primeiro plano das
explica\c{c}\~{o}es cient\'{\i}ficas.

         Al\'em de Fourier, tamb\'em Lagrange teve um papel
fundamental na afirma\c{c}\~{a}o desse sistema explicativo (MOREIRA XAVIER,
1986). Ao utilizarmos as equa\-\c{c}\~{o}es de Lagrange (obtidas a partir do
chamado {\it princ\'{\i}pio de m\'{\i}nima a\c{c}\~{a}o}), para resolver um problema
espec\'{\i}fico e explicar um fen\^omeno (LANCZOS, 1986), estamos
atribuindo a ele, al\'em da {\it causa formalis} dada pela
lagrangeana, uma {\it causa finalis} expressa pelo princ\'{\i}pio
variacional. Foi o estudo de sistemas complexos -- propaga\c{c}\~{a}o de
calor, mec\^anica dos fluidos e teoria dos campos -- que exigiu o
uso de um sistema explicativo complexo (quatro causas) e o abandono
do mecanicismo {\it stricto sensu}, baseado exclusivamente na {\it
causa efficiens} (MOREIRA XAVIER, 1993).

          Ainda no interior da F\'{\i}sica Cl\'assica, o Eletromagetismo oferece,
tamb\'em, um exemplo interessante. Embora as equa\c{c}\~{o}es de Maxwell tenham
sido obtidas a partir de um modelo mec\^anico do \'eter -- no fundo,
portanto, de um esquema baseado na {\it causa efficiens} --
Hertz percebeu que este esquema tinha que ser abandonado, em
benef\'{\i}cio da {\it causa formalis}, ao nos ensinar que as Equa\c{c}\~{o}es de
Maxwell {\it s\~ao} a Teoria de Maxwell. Em suas palavras:

\begin{quotation}
{\noindent \it ``O que \'e a teoria de Maxwell?" N\~ao conhe\c{c}o resposta
mais sucinta nem mais definitiva do que a seguinte: -- a teoria de Maxwell
\'e o sistema de equa\c{c}\~{o}es de Maxwell.''} (HERTZ, 1893).
\end{quotation}

            A F\'{\i}sica Contempor\^anea est\'a repleta de exemplos de
utiliza\c{c}\~{a}o da {\it causa formalis}. V\'arias propriedades qu\^anticas da
mat\'eria, como spin, paridade, isospin, estranheza, char\-me, cor dos
quarks, dentre outras, s\~ao pass\'{\i}veis de descri\c{c}\~{a}o apenas em termos
matem\'a\-ti\-cos.  A rela\c{c}\~{a}o entre os teoremas de conserva\c{c}\~{a}o e as
propriedades de simetria \'e outro exemplo: quando se diz que uma grandeza
se conserva, como resultado de uma determinada simetria, o que se est\'a a
fazer \'e atribuir a conserva\c{c}\~{a}o a uma {\it causa formalis}.  Na teoria
da Relatividade de Einstein, o programa de geometriza\c{c}\~{a}o da F\'{\i}sica \'e
claramente calcado na primazia dos aspectos formais, ou seja, na
valoriza\c{c}\~{a}o da {\it causa formalis}.

              Existe um exemplo interessante em Teoria de Campos onde
se pode mostrar que {\it causa formalis et causa efficiens}
coe\-xistem num esquema explicativo mais complexo: \'e o processo
din\^amico de quebra espont\^anea de simetria, onde a densidade de
{\it lagrangeana} apresenta explicitamente uma certa {\it simetria}
-- {\it causa formalis} -- que n\~ao mais se manifesta nas
equa\c{c}\~{o}es de movimento devido a alguma {\it particularidade das
intera\c{c}\~{o}es} entre os campos que comp\~oem a densidade de
lagrangeana --- a {\it causa efficiens} da quebra de simetria (que
n\~ao \'e obviamente devida a uma for\c{c}a no sentido newtoniano).

              Por \'ultimo, o {\it princ\'{\i}pio de exclus\~ao} de
Pauli, que vale para um s\'o tipo de mat\'eria -- de \textit{spin}
semi-inteiro, {\it i.e.}, os f\'ermions --- e n\~ao para part\'{\i}culas
de spin inteiro -- os b\'osons ---  pode estar, de alguma forma,
relacionado \`a {\it causa materialis} (MOREIRA XAVIER, 1986). Usar o
princ\'{\i}pio de Pauli para explicar a impenetrabilidade da mat\'eria
significa invocar uma {\it causa materialis} para este fen\^omeno.

          Esta lista de exemplos, ainda que bastante incompleta, serve,
entretanto, para ilustrar como as explica\c{c}\~{o}es de que a F\'{\i}sica lan\c{c}a
m\~ao  hoje em dia s\~ao mais variadas (e, em geral, mais complexas), do
que as concebidas nos estreitos limites do programa mecanicista, na medida
em que incorporam todos os quatro tipos de causa.

Gostar\'{\i}amos de concluir esta Se\c{c}\~{a}o reafirmando que, no diversificado
sistema explicativo  da F\'{\i}sica Contempor\^anea, {\it causa} readquire um
sentido lato (KUHN, 1977), cujas ra\'{\i}zes remotas podem ser encontradas no
esquema aristot\'elico-escol\'astico de quatro causas. Este fato n\~ao
chega a surpreender se nos lembrarmos, por exemplo, do car\'ater
neo--aristot\'elico da F\'{\i}sica de Einstein, assinalado por Koyr\'e
(KOYR\'E, 1971, p. 269).

\section{Arist\'{o}teles}
\bigskip

          Que a tridimensionalidade \'e para Arist\'oteles um atributo do
corp\'oreo -- suposto completo em magnitude e imut\'avel -- se aprende
no primeiro cap\'{\i}tulo do seu {\it De Caelo}:
\begin{quotation}
 {\noindent \it Se uma magnitude \'e divis\'{\i}vel de um modo, \'e uma
linha, se de dois modos, \'e uma superf\'{\i}cie e se de tr\^es, um
corpo. Al\'em dessas, n\~ao h\'a outra magnitude, porque tr\^es s\~ao
todas as dimens\~oes que existem, e o que \'e divis\'{\i}vel em tr\^es
dire\c{c}\~{o}es \'e divis\'{\i}vel em todas. (...) posto que `todas' e `tudo'
e `completo'} [s\~ao conceitos que] {\it n\~ao diferem entre si no
que diz respeito \`a forma, mas apenas, quando muito, diferem nas
suas mat\'erias e naquilo a que eles s\~ao a\-pli\-ca\-dos, \underbar
{s\'o o corpo, entre as magnitudes, pode ser completo}. \underbar
{Pois}, \underbar {so\-men\-te} \underbar {ele \'e determinado por
tr\^es dimens\~oes}, isto \'e, em um `todo' (...) N\'os n\~ao podemos
passar do corpo para outra coisa, como passamos da linha para a
superf\'{\i}cie, e da superf\'{\i}cie para o corpo. Pois, se pud\'essemos,
n\~ao seria mais verdade que o corpo \'e magnitude completa.
Poder\'{\i}amos passar al\'em dele apenas em virtude de um defeito nele
existente, e o que \'e completo n\~ao pode ser deficiente, pois se
estende em todas as dire\c{c}\~{o}es.''} (tradu\c{c}\~{a}o e grifos dos
autores)\footnote{A magnitude if divisible one way is a line, if
two ways a surface, and if three a body. Beyond these there is no other
magnitude, because the three dimensions are all that there are, and that
which is divisible in three directions is divisible in all. (...) since
`every' and `all' and `complete' do not differ from one another in respect
of form, but only, if at all, in their matter and in that to which they are
applied, body alone among magnitudes can be complete. For it alone is
determined by the three dimensions, that is, in an `all' (...) We cannot
pass beyond body to a further kind, as we passed from length to surface,
and from surface to body. For if we could, it would cease to be true that
body is complete magnitude. We could pass beyond it only in virtue of a
defect in it and that which is complete cannot be defective, since it
extends in every directions'' (ARIST\'OTELES, {\it in} BARNES,
1985).}
\end{quotation}

          Portanto, da leitura de Arist\'oteles, concluimos que o {\it Ser}
-- o corp\'oreo em sua completude -- \'e a {\it causa materialis} da
tridimensionalidade, negada ao {\it topos}, que \'e uma extens\~ao
bidimensional, como notou {\it Simplicius} (SIMPLICIUS, {\it apud} JAMMER,
1954).

\section{Kant}

\'E difundida na literatura a opini\~ao de que Kant apresentou a
primeira solu\c{c}\~{a}o f\'{\i}sica para a quest\~ao da dimensionalidade (JAMMER,
1954; BRITTAN, 1978; BARROW, 1983; BARROW \& TIPLER, 1986; CARUSO \&
MOREIRA XA\-VIER, 1987).  A ele \'e atribu\'{\i}do o argumento de que a {\it
raz\~ao da tridimensionalidade do espa\c{c}o poderia ser encontrada na lei
da gravita\c{c}\~{a}o de Newton, segundo a qual a for\c{c}a entre dois corpos
decresce com o quadrado da dist\^ancia que os separa}. Tal concep\c{c}\~{a}o se
ap\'oia, provavelmente, no t\'{\i}tulo do d\'ecimo par\'agrafo do escrito {\it
Considera\c{c}\~{o}es Sobre a Verdadeira Estimativa das For\c{c}as Vivas}, de
Kant, a saber:
\begin{quotation}
{\noindent \it ``\'E prov\'avel que a tridimensionalidade seja devida \`a
lei que define as for\c{c}as que as subst\^ancias exercem umas sobre as
outras}" (KANT, 1747, p.~11).
\end{quotation}

Entretanto, veremos nesta Se\c{c}\~{a}o, que uma leitura mais
cuidadosa da obra acima citada, como um todo, leva-nos a concluir que
seu racioc\'{\i}\-nio n\~ao conduz a uma resposta satisfat\'oria sobre a
dimensionalidade do {\it espa\c{c}o} -- como se insinua no t\'{\i}tulo e \'e
normalmente aceito -- mas limita-se, na verdade, a justificar a
tridimensionalidade da {\it extens\~ao}. Esta quest\~ao ser\'a examinada em
outro trabalho (CARUSO \& MOREIRA XAVIER, 1994).

Aqui cabe, apenas, mencionar dois pontos: em primeiro
lugar, que essa ideia de Kant --- elaborada no per\'{\i}odo
pr\'e--cr\'{\i}tico --- de procurar determinar a dimensiona\-li\-dade do
espa\c{c}o  a partir de uma lei f\'{\i}sica \'e, sem d\'uvida, um marco
important\'{\i}ssimo para a discuss\~ao moderna deste problema, embora
n\~ao se sustente no per\'{\i}odo cr\'{\i}tico da filosofia
kantiana;\footnote{Esta
delicada quest\~ao \'e tratada em (VUILLEMIN, 1967) e discutida pelos
autores (CARUSO \& MOREIRA XAVIER, 1994).} em segundo lugar, que
h\'a um ponto pac\'{\i}fico, qualquer que seja a leitura que se fa\c{c}a
desses textos pr\'e--cr\'{\i}ticos de Kant: a grande import\^ancia da lei
de Newton em seu argumento, ou em outras palavras o papel fundamental
que a {\it for\c{c}a} desempenha em sua explica\c{c}\~{a}o. Esta afirmativa
\'e corroborada pelas cita\c{c}\~{o}es: ``\textit{Todo corpo tem uma for\c{c}a que lhe \'e essencial.}'' (KANT, 1747, p.~3).

\begin{quotation}
{\noindent \it ``Prova--se facilmente que n\~ao haveria espa\c{c}o nem
extens\~ao se as subst\^ancias n\~ao tivessem for\c{c}as pelas quais
atuassem fora de seus limites. Pois sem uma for\c{c}a deste tipo n\~ao h\'a
conex\~{a}o, sem esta conex\~{a}o n\~ao h\'a ordem e sem esta ordem
n\~ao h\'a espa\c{c}o.}'' (KANT, 1747, p.~10);
\end{quotation}

\begin{quotation}{\noindent \it ``Posto que devemos ser capazes de deduzir tudo que
se encontre entre as qualidades de uma coisa a partir daquela que contenha
em si o fundamento mais completo da pr\'opria coisa, as qualidades da
extens\~ao, e consequentemente sua tridimensionalidade, fundamentar-se-\~ao
nas qualidades da for\c{c}a que as subst\^ancias possuem na presen\c{c}a das
coisas com as quais elas est\~ao relacionadas.}'' (KANT, 1747, p.~11).
\end{quotation}

    A for\c{c}a \`a qual Kant se refere nestes textos \'e dada pela lei de
atra\c{c}\~{a}o de Newton que depende do inverso do quadrado das dist\^ancias.

          Este fato indica que a {\it causa efficiens} tem um papel
fundamental na explica\c{c}\~{a}o que Kant apresenta para a tridimensionalidade e
pode ser considerado uma indica\c{c}\~{a}o de quanto ele n\~ao compartilha das
ideias galileanas de geometriza\c{c}\~{a}o da F\'{\i}sica.  Seu sistema explicativo
\'e essencialmente newtoniano, e toda a sua argumenta\c{c}\~{a}o \'e constru\'{\i}da a
partir das leis de for\c{c}a. \'E como se ele entendesse que o estudo de
espa\c{c}os mais gen\'ericos devesse preceder a discuss\~ao da
dimensionalidade do espa\c{c}o no \^ambito da F\'{\i}sica, antecipando uma
interrela\c{c}\~{a}o dos dois problemas. Mesmo que sua conjectura refira-se \`as
dimens\~oes da {\it extens\~ao}, Kant foi obrigado a imaginar a
possibilidade de exist\^encia de espa\c{c}os com um n\'umero diferente de
dimens\~oes, antes que houvesse uma teoria para estes tipos de espa\c{c}o.
Ser\'a a descoberta das geometrias n\~ao-euclidianas, no s\'eculo XIX, que
dar\'a impulso a estas quest\~oes (JAMMER, 1954).  Parece-nos, portanto,
que Kant n\~ao s\'o tinha consci\^encia, j\'a em 1747, da dualidade a que
se referiu Whitrow (WHITROW, 1955, pp. 13-31), mas, sobretudo, lan\c{c}ou as
suas bases.

          Por outro lado, Kant rompe tamb\'em com a concep\c{c}\~{a}o
aristot\'elica do problema -- no seu aspecto geral (causa do espa\c{c}o) e
particular (causa da dimensionalidade) -- atrav\'es da introdu\c{c}\~{a}o da
{\it for\c{c}a} como {\it causa efficiens} do espa\c{c}o, via o conceito de
{\it ordem}\footnote{O fato de Kant
considerar a for\c{c}a como essencial ao corpo sugere que esta possa ter
ainda, para ele, um certo car\'ater de ess\^encia (ou forma),
reminisc\^encia talvez aristot\'elica, que nos permitiria imaginar a
for\c{c}a tamb\'em como uma esp\'ecie de \textit{causa
formalis.} Embora aristot\'elico no papel que a {\it subst\^ancia} desempenha em seu sistema explicativo, note que Kant
considera, aqui,  {\it for\c{c}a} como geradora de ordem (KANT, 1747, p.
10), ao contr\'ario de Arist\'oteles, em cujo sistema, {\it for\c{c}a
(dynamis)} conduz \`a ruptura de ordem c\'osmica.}

          Esta dupla ruptura -- com Galileu e Arist\'oteles -- demonstra
a originalidade de Kant e, no fundo, aponta para a grande import\^ancia que
sua ideia ter\'a depois da inven\c{c}\~{a}o dos conceitos de linha de for\c{c}a
e de campo. De fato, do ponto de vista da F\'{\i}sica, uma compreens\~ao mais
profunda da conjectura de Kant s\'o pode ser alcan\c{c}ada com o conceito de
campo, com suas implica\c{c}\~{o}es j\'a mencionadas na Se\c{c}\~{a}o III. \'E
atrav\'es da solu\c{c}\~{a}o da equa\c{c}\~{a}o de Laplace-Poisson em um espa\c{c}o
euclidiano de $n$-dimens\~oes (ou, equivalentemente, pela aplica\c{c}\~{a}o do
teorema de Gauss ao campo gravitacional produzido por uma massa) que se
p\~oe em evid\^encia a rela\c{c}\~{a}o entre o expoente do potencial newtoniano e
a dimensionalidade do espa\c{c}o.

          At\'e onde sabemos -- com o aval de Brittan (BRITTAN, 1978, pp.
96-97) --, n\~ao h\'a outra tentativa de Kant para fornecer uma base f\'{\i}sica
\`a quest\~ao da dimensio\-nalidade. Sabe--se que Kant voltou a este
problema, como atestam os manuscritos coligidos no {\it Opus Postumum}
(KANT, 1986), mas, ironicamente, h\'a uma interrup\c{c}\~{a}o no texto, num ponto
fundamental, tornando imposs\'{\i}vel descobrir como o Kant maduro revisitaria
o problema da dimensionalidade do espa\c{c}o. Concluiremos esta Se\c{c}\~{a}o com
esta reticente cita\c{c}\~{a}o de Kant:
\begin{quotation}
{\noindent {\it ``A qualidade do espa\c{c}o e do tempo, por exemplo que o primeiro tenha tr\^es dimens\~{o}es, o segundo somente uma, que a revolu\c{c}\~{a}o seja regida pelo quadrado das dist\^ancias s\~ao princ\'{\i}pios que (...)} [interrup\c{c}\~{a}o]}'' (KANT, 1986, p. 131).
\end{quotation}

\section{Ehrenfest}
\medskip
          Em seu artigo sobre a dimensionalidade do espa\c{c}o, Ehrenfest
introduz um novo enfoque para o problema, dando-lhe uma base f\'{\i}sica mais
s\'olida (EHRENFEST, 1920, p. 440). Considera, como ponto de partida, a
hip\'otese de que a din\^amica de um sistema planet\'ario em um espa\c{c}o
de $n$--dimens\~oes, seja regida pelo potencial solu\c{c}\~{a}o da equa\c{c}\~{a}o de
Laplace--Poisson neste espa\c{c}o, obtida por uma extens\~ao formal do
operador laplaciano de tr\^es a $n$-dimens\~oes. Desse modo, mostra que o
sistema planet\'ario s\'o admite solu\c{c}\~{o}es mecanicamente est\'aveis para
$n=3$. Em suma, Ehrenfest busca identificar diferen\c{c}as entre alguns
fen\^omenos f\'{\i}sicos em um espa\c{c}o tridimensional ($\Re^3$) e noutro
n-dimensional ($\Re^n$). A estes aspectos que distinguem a F\'{\i}sica em
$\Re^3$ daquela em $\Re^n$ ele d\'a o nome de {\it aspectos singulares}, e
seu trabalho \'e dedicado a p\^o-los em evid\^encia. O ponto central da
abordagem de Ehrenfest consiste nas seguintes hip\'oteses: {\it (i)} \'e
poss\'{\i}vel fazer uma extens\~ao formal de certas leis f\'{\i}sicas estabelecidas
no $\Re^3$ para o $\Re^n$; {\it (ii)} a partir da\'{\i}, podem--se buscar um ou
mais princ\'{\i}pios -- por exemplo, o princ\'{\i}pio da estabilidade -- que,
junto com esta lei, permitir\~ao determinar o n\'umero de dimens\~oes do
espa\c{c}o f\'{\i}sico.  Este tipo de abordagem foi seguido por outros autores,
que mostraram que o \'atomo hidrogen\'oide de Bohr--Schr\"odinger e a
solu\c{c}\~{a}o de Schwarzschild das equa\c{c}\~{o}es de Einstein apresentam tamb\'em
aspectos singulares para $n=3$ (WHITROW, 1959; TANGHERLINI, 1963 e 1986).
Uma cr\'{\i}tica epistemol\'ogica \`a ess\^encia destes trabalhos foi apresentada
pelos autores (CARUSO \& MOREIRA XAVIER, 1987).

          Gostar\'{\i}amos de enfatizar que, para a ideia de Ehrenfest ser
implementada, em geral \'e a {\it estrutura matem\'atica} de certo
formalismo f\'{\i}sico, ou simplesmente a {\it forma} de uma equa\c{c}\~{a}o
f\'{\i}sica,\footnote{N\~ao \'e demais repetir,
estabelecida pressupondo a tridimensionalidade.} que \'e mantida, e sua
validade em um espa\c{c}o com um n\'umero arbitr\'ario de dimens\~{o}es, {\it
postulada}. Isto revela o papel da {\it causa formalis} na resposta
que d\'a Ehrenfest \`a quest\~ao: por que o espa\c{c}o tem tr\^es
dimens\~oes? Por\'em h\'a mais ingredientes no argumento de Ehrenfest: \'e
preciso invocar tamb\'em a {\it causa efficiens}  para se entender a
instabilidade dos sistemas din\^amicos consi\-de\-rados (planet\'ario e
hidrogen\'oide)  para $n\not= 3$. Este fato, mais uma vez ilustra a
complexidade das modernas explica\c{c}\~{o}es causais.
\bigskip

\section{Considera\c{c}\~{o}es Finais e Conclus\~{o}es}

          A dimensionalidade do espa\c{c}o vem sendo problematizada desde
Plat\~ao e Aris\-t\'oteles, mas com grande descontinuidade (WEYL, 1949;
JAMMER, 1954; REICHENBACH, 1957; GR\"UNBAUM, 1974; BARROW \& TIPLER, 1986;
CA\-RU\-SO \& MOREIRA XAVIER, 1987).  No s\'eculo XIX, com o sur\-gi\-mento da
geo\-metria n\~ao-euclidiana, este assunto come\c{c}a a deixar de ser objeto
de espe\-cula\c{c}\~{a}o filos\'ofica e passa a ser considerado um problema
f\'{\i}sico ou l\'ogico (JAMMER, 1954). Nas primeiras d\'ecadas do s\'eculo XX,
sem d\'uvida, \'e a Teoria da Relatividade que vai reacender a discuss\~ao
em torno deste tema (REICHENBACH, 1957; GR\"UNBAUM, 1974).  Nos \'ultimos
anos, esta quest\~ao vem adquirindo novo interesse na F\'{\i}sica devido,
principalmente, ao desenvolvimento de dois programas, a saber: tentativas
de unifica\c{c}\~{a}o das quatro intera\c{c}\~{o}es fundamentais atrav\'es de teorias
cons\-tru\-\'{\i}das em espa\c{c}os com maior n\'umero de dimens\~oes, como
supergravidade e {\it superstrings} ({\it Cf.}, por ex., SRIVASTAVA, 1986;
DAVIES \& BROWN, 1988), e o estudo de sistemas n\~ao lineares envolvendo
estruturas fractais (MANDELBROT, 1987). Portanto, a quest\~ao da
dimensionalidade do espa\c{c}o est\'a longe de ser uma quest\~ao fechada, ou
meramente acad\^emica, e \'e ainda objeto de intensa investiga\c{c}\~{a}o
cient\'{\i}fica.

          Al\'em disso, existem numerosas quest\~oes epistemol\'ogicas
referentes \`a dimensio\-nalidade do espa\c{c}o, que precisam ser elucidadas.
Neste ensaio, discutiu--se a rela\c{c}\~{a}o entre os crit\'erios utilizados para
explicar ou impor limites sobre a dimensionalidade do espa\c{c}o e o sistema
de explica\c{c}\~{a}o causal dominante no correspondente per\'{\i}odo hist\'orico,
focalizando-se as contribui\c{c}\~{o}es marcantes de Arist\'oteles, Kant e
Ehrenfest ao problema da di\-men\-sio\-na\-lidade. Vimos que os argumentos
(e consequentes limita\c{c}\~{o}es) dos tr\^es sedimentam-se em explica\c{c}\~{o}es
causais distintas: em Arist\'oteles a tridimensionalidade \'e um atributo
das subst\^ancias corp\'oreas ({\it causa materialis}); Kant parte de uma
concep\c{c}\~{a}o newtoniana de for\c{c}a, que se torna basicamente a {\it causa
efficiens} da tridimensionalidade (KANT, 1747), enquanto Ehrenfest admite,
como ponto de partida de sua argumenta\c{c}\~{a}o (na passagem $\Re^3\rightarrow
\Re^n$) (EHRENFEST, 1920, p.~440), um referencial te\'orico constru\'{\i}do,
principalmente, a partir da ideia de {\it causa formalis}.

Excetuando-se alguns argumentos de natureza topol\'ogica
(fundamentados  na teoria dos n\'os, por exemplo), n\~ao encontramos
na literatura especializada recente nenhum procedimento que n\~ao se
baseie explicitamente em {\it equa\c{c}\~{o}es diferenciais}.  E mais: toda
vez que se consegue obter v\'{\i}nculos sobre a dimensionalidade
topol\'ogica do espa\c{c}o, as equa\c{c}\~{o}es em quest\~ao s\~ao lineares
(n\~ao importando o m\'etodo utilizado). Alguns e\-xemplos s\~ao as
equa\c{c}\~{o}es de Laplace-Poisson (EHRENFEST, 1920, p.~440), de
d'Alembert (HADAMARD, 1923; B\"UCHEL, 1963), de Schr\"odinger
(TANGHERLINI, 1963; CARUSO \& MOREIRA XAVIER, 1987), de Klein--Gordon
(CARUSO, NETO, SVAITER \& SVAI\-TER, 1991) e de Dirac (CARUSO \&
MO\-REIRA XAVIER, 1987; L\"AMMER\-ZAHL, 1993). Espera-se que
equa\c{c}\~{o}es n\~ao-lineares d\^eem v\'{\i}nculos sobre a dimensionalidade
fractal do espa\c{c}o. Al\'em disso, \'e de se supor que as teorias de
campo unificadas sejam capazes de iluminar certos aspectos da
quest\~ao da dimensio\-na\-lidade do espa\c{c}o f\'{\i}sico. Se esta
perspectiva se confirmar, mais uma vez verificaremos que a
Matem\'atica desempenha um papel fundamental na tentativa de se
compreender esta quest\~ao, n\~ao como ci\^encia pura, mas como
\'unico modo aceito, desde a re\-vo\-lu\-\c{c}\~{a}o galileana, de
formalizar um certo conhecimento do mundo fenomenol\'ogico.  Este
fato aponta para uma crescente valoriza\c{c}\~{a}o da {\it causa formalis}
na abordagem do problema da dimensionalidade do espa\c{c}o f\'{\i}sico.
\vskip 1.0 true cm

\noindent {\bf Agradecimento:}  Um de n\'os (F.C.) agradece ao CNPq, pela
bolsa de pesquisa vigente durante o per\'{\i}odo no qual este trabalho foi
realizado.

\vspace*{1cm}
\section{REFER\^ENCIAS}

\begin{itemize}

\item ARIST\'OTELES, {\it De Caelo,} Livro I, 268a, 5 e sq., {\it
in} BARNES, J. (ed.) {\it The Complete Works of Aristotle}. Princeton:
Princeton Univ. Press, 1985.

\item  BARROS, J.A. de. {\it Dois exemplos de indecidibilidade e
incompletude em F\'{\i}sica}, Tese de Doutorado. Rio de Janeiro: CBPF, 1991.

\item   BARROW, J.D. ``Dimensionality", {\it Phil. Trans. Roy. Soc.
London} v. A310, p. 337, 1983.

\item  BARROW, J.D. \& TIPLER, F.J. {\it The Anthropic
Cosmological Principle}. Oxford: Claredon Press, 1986.

\item  BOIS-REYMOND, E du. {\it \"Uber die Grenzen des
Naturerkennens}. Leipzig: Verlag von Veit \& Comp., 1891; tradu\c{c}\~{a}o
italiana aos cuidados de CAPPELLETTI, V. {\it I confini della conoscenza
della natura}. Milano: Feltrinelli Editore, 1973.

\item   BRITTAN Jr., G.G. {\it Kant's Theory of Science}.
Princeton: Princeton Univ. Press, 1978.

\item  B\"UCHEL. W. ``Warum hat der Raum drei Dimensionen?", {\it
Physikalische Bl\"atter} v. 19, pp. 547-49, 1963; traduzido e adaptado por
FREEMAN, I.M. com o t\'{\i}tulo ``Why is Space Three-Dimensional", {\it
American Journal of Physics} v. 37, p. 1222, 1969.

\item  BUNGE, M. {\it Causality and Modern Science,} $3^{\a}$ ed.
revisada. New York: Dover, 1979.

\item  CARUSO, F. \& MOREIRA XAVIER, R. ``On the physical problem
of spatial dimensions: an alternative procedure to stability arguments",
{\it Fundamenta Scientiae,} v. 8. n. 1, pp. 73-91, 1987.

\item  ---. ``Notas sobre o problema da dimensionalidade nos
primeiros textos do jovem Kant", {\it Notas de F\'{\i}sica} CBPF-NF-050/94.
Rio de Janeiro: Centro Brasileiro de Pesquisas F\'{\i}sicas, 1994.

\item  CARUSO, F., NETO, N.P., SVAITER, B. \& SVAITER, N.
``Attractive or Repulsive Nature of Casimir Force in D-Dimensional
Minkowski Spacetime", {\it Physical Review D} v. 43, n. 4, pp. 1300-6,
1991.

\item  COSTA, N.C.A. da \& DORIA, F.A. ``Undecidability and
incompleteness in Classical Mechanics", {\it Int. J. Theor. Phys.} v. 30,
p. 1041 (1991).

\item  DAVIES, P.C.W. \& BROWN, J. (Eds.), {\it Superstrings: a
theory of everything?}. Cambridge: Cambridge Univ. Press, 1988.

\item  DIJKSTERHUIS, E.J. {\it De Mechanisering van het
Wereldbeeld}. Amsterdam: 1959; tradu\c{c}\~{a}o inglesa de DIKSOORN, C. {\it The
Mechanization of the World Picture: Pytha\-goras to Newton}. Princeton:
Princeton Univ. Press, 1986.

\item  EHRENFEST, P. ``Welche Rolle spielt die
Dreidimensionalit\"at des Raumes in den Grund\-gesetzen der Physik?", {\it
Ann.  Physik} v.  61, p. 440, 1920. {\it Cf.} tamb\'em seu ``In what way
does it become manifest in the fundamental laws of physics that space has
three dimensions?", {\it Proc. Amsterdam Acad.} v. 20, p. 200, 1917.
(Reimpresso em KLEIN, M.J. (ed.) {\it Paul Ehrenfest -- Collected
Scientific Papers}. Amsterdam: North Holland Publ. Co., 1959. pp. 400-409).

\item  FOURIER, J.B.J. {\it La Th\'eorie Analytique de la Chaleur},
1822, {in \OE uvres} pu\-blicadas por DARBOUX, G. Paris: \'edition
Gauthier--Villars, t. I, 1888, t. II 1890. (Tradu\c{c}\~{a}o inglesa {\it in}
vol.  45, {\it Great Books}. Chicago: Enc. Britannica, 1952. p. 169).

\item  GR\"UNBAUM, A. {\it Philosophical Problems of Space and
Time}, second, enlarged edition. Dordrecht: D. Reidel Publ., 1974.

\item  HADAMARD, J. {\it Lectures on Cauchy's problem in linear
partial differential equations}. New Haven: Yale Univ. Press, 1923. pp.
53-54, 175-177 e 235-236.

\item  HERTZ, H. {\it Electric Waves -- being Researches on the
Propagation of Electric Action with Finite Velocity through Space}.
Macmillan and Co., 1893. (Reeditado em New York: Do\-ver, 1962, p. 21).

\item  JAMMER, M. {\it Concepts of Space: the History of Theories
of Space in Physics}. Cambridge: Harvard Univ. Press,  1954.

\item  ---. {\it Concepts of Force: a Study in the Foundations of
Dynamics}. Cambridge: Harvard Univ. Press, 1957. (Tra\-du\-\c{c}\~{a}o italiana
de BELLONE, E. {\it Storia del Concetto do Forza: Studio sulle Fondazioni
della Dinamica}. Milano: Feltrinelli Editore, seconda edizione, 1979).

\item  KANT, I. {\it Gedanken von der wahren Sch\"atzung der
lebendigen Kr\"afte}. K\"onigs\-berg: 1747. (Tradu\c{c}\~{a}o inglesa de
HANDYSIDE, J. {\it in Kant's inaugural dissertation and the early writings
on space}. Chicago: Open Court, 1929).

\item  ---. {\it Opus Postumum -- passage des principes
m\'etaphysiques de la science de la nature \`a la physique}, tradu\c{c}\~{a}o,
apresenta\c{c}\~{a}o e notas de MARTY, F. Paris: Presses Universitaires de
France, 1986.

\item  KOYR\'E, A. {\it Etudes galil\'eennes}. Paris: Hermann, 1966.

\item  ---. {\it \'Etudes d'histoire de la pens\'ee
phylosophique}. Paris: Gallimard, 1971.

\item  KUHN, T. {\it The Essential Tension: Selected Studies in
Scientific Tradition and Change}. Chicago: Univ. Chicago, 1977.
(Tradu\c{c}\~{a}o portuguesa de PACHECO, R. {\it A Tens\~ao Essencial},
Lisboa, Edi\c{c}\~{o}es 70, s/d. {\it Cf.} o artigo ``Conceitos de causa no
desenvolvimento da F\'{\i}sica", pp. 51-61).

\item  L\"AMMERZAHL, C. \& MACIAS, A. ``On the dimensionality of
space--time", {\it Journal of Mathematical Physics} v. 34, 4540, 1993.

\item  C. LANCZOS, C. {\it The Variational Principles of Mechanics},
quarta edi\c{c}\~{a}o.  New York: Dover, 1986.

\item  LAPLACE, P.S.  {\it Teoria Anal\'{\i}tica das Probabilidades
--- Pref\'acio}, 1814. {\it apud} FERRATER MORA, J. {\it Diccionario de
Filosofia}. Madrid: Alianza, 1982.

\item  MANDELBROT, B.B. {\it Les Objets Fractals}; edi\c{c}\~{a}o italiana
aos cuidados de R. Pigno\-ni, {\it Gli Oggetti Frattali: Forma, Caso e
Dimensione}. Torino: G. Einaudi, 1987.

\item  MOREIRA XAVIER, R. ``Notas sobre a evolu\c{c}\~{a}o do conceito de
causa na f\'{\i}sica p\'os-newtoniana: da causa eficiente \`a causa
formal". {\it Notas de F\'{\i}sica} CBPF-NF-053/86.
Rio de Janeiro: Centro Brasileiro de Pesquisas F\'{\i}sicas, 1986.

\item  ---.  ``Bachelard e o Livro do Calor: o Nascimento da F\'{\i}sica
Mate\-m\'a\-tica na \'Epoca da Articula\c{c}\~{a}o Causal do Mundo", {\it Revista
Filos\'ofica Brasileira} v. 6, n. 1, pp. 100-113, 1993.

\item  NEWTON, I. {\it Opticks or A Treatise of the Reflections,
Refractions, Inflections \& Colors of Light} (baseada na quarta edi\c{c}\~{a}o,
Londres, 1730). New York: Dover, 1952, {\it Query} 31.

\item  ---. {\it Philosophi\ae Naturalis Principia Mathematica}.
The third editon (1726) with variant readings assembled and edited by
KOYR\'E, A. and  COHEN, I.B. Cambridge: Harvard University Press,
1972, {\it Rule} 3 {\it Book} 3.

\item  REICHENBACH, H. {\it Philosophie der Raum-Zeit-Lehre}.
(Tradu\c{c}\~{a}o inglesa de REICHENBACH, M. \& FREUND, J. {\it The Philosophy of
Space \& Time}. New York: Dover, 1957, pp. 273-83).

\item  SCHENBERG, M. {\it Pensando a F\'{\i}sica}. S\~ao Paulo: Ed.
Brasiliense, $2^{\a}$ Ed, 1985.

\item  SIMPLICIUS, {\it F\'{\i}sica}, 601, {\it apud} (JAMMER, 1954).

\item  SRIVASTAVA, P.P. {\it Supersymmetry, Superfields and
Supergravity: an Introduction}. Bristol: Adam Hilger, 1986.

\item  TANGHERLINI, F.R. ``Schwarzschild Field in n-Dimensions
and the Dimensionality of Space Problem", {\it in} {\it Nuovo Cimento}
v. 27, p. 636, 1963.

\item  ---. ``Dimensionality of Space and the Pulsating Universe",
{\it ibid} v. 91B, p. 209, 1986.

\item  VUILLEMIN, J. ``La th\'eorie kantienne de l'espace \`a la
lumi\`ere de la th\'eorie des groupes de transformations", {\it The
Monist}, v. 51, No. 3, pp. 332-351, 1967; reimpresso {\it In} {\it
L'Intuitionnisme Kantien}. Paris: Vrin, 1994.

\item  WEYL, H. {\it Philosophy of Mathematics and Natural
Science.} Princeton: Princeton University Press, 1949.

\item  WHITROW, G.J. ``Why Physical Space has three Dimensions",
{\it Brit. J. Phil. Sci.} v. 6, pp. 13-31, 1955.

\item  ---. {\it The Structure and Evolution of the Universe}. New
York: Harper and Row, 1959; {\it Cf.} tamb\'em (WHITROW, 1955, pp. 13-31).
\end{itemize}

\end{document}